\documentclass
[rmp,twocolumn,amsmath,letterpaper,raggedbottom,nobalancelastpage]{revtex4}

\usepackage[english,german]{babel}
\usepackage{ifpdf}
\ifpdf
  \def\hrefx#1#2#3{\href{#1}{{#2}\ {#3}}}
\else
  \def\hrefx#1#2#3{\href{#1}{#2}\penalty0\ \href{#1}{#3}}
\fi

\usepackage{times}
\usepackage[bookmarks=false]{hyperref}
\usepackage{soul}
\sodef\spaceout{}{0.1em}{0.3em plus 0.2em}{0.3em plus 0.2em}
\usepackage{array}

\hypersetup{pdftitle={\"Uber die Berechnung der geographischen L\"angen
und Breiten aus geod\"atischen Vermessungen},
pdfauthor={F. W. Bessel},
pdfsubject={Astron. Nachr. 4(86), 241-254 (1825)},
pdfkeywords={doi:10.1002/asna.18260041601}}

\makeatletter\def\raggedcolumn@skip{\vskip\z@\@plus.0001fil\relax}\makeatother

\newcommand{\p}{\!\cdot\!}
\def\dms#1:#2:#3:{#1^\circ\,#2'\,#3''}
\newcommand{\colog}{\mathop{\mathrm{colog}}\nolimits}
\def\sumbar#1{\overline{\vphantom{\rule{0pt}{2.5ex}}#1}}
\def\b#1{\ifcase#1\relax0\or9\or8\or7\or6\or5\or4\or3\or2\or1\fi}

\def\/{\nobreak\hspace{0.08em}}
\date[Originally published: ]{October 1825;
transcribed: August 13, 2009; revised: March 29, 2012}

\begin{document}

\title{"Uber die Berechnung der geographischen L"angen und Breiten aus
geod"atischen Vermessungen\footnote
{\selectlanguage{english}This transcription of
\href{http://adsabs.harvard.edu/full/1825AN......4..241B}%
{Astronomische Nachrichten {\bf 4}(86), 241--254 (1825)},
doi:\href{http://dx.doi.org/10.1002/asna.18260041601}
{10.1002/asna.18260041601},
has been edited by Charles F. F. Karney
$\langle$\href{mailto:charles.karney@sri.com}{charles.karney@sri.com}$\rangle$
and Rodney E. Deakin
$\langle$\href{mailto:rod.deakin@rmit.edu.au}{rod.deakin@rmit.edu.au}$\rangle$.
The paper also appears in
\hrefx{http://books.google.com/books?id=vX4EAAAAYAAJ&pg=PA5}
{{\it Abhandlungen von Friedrich Wilhelm Bessel},}{Vol.~3}
(W. Engelmann, Leipzig, 1876).
The text follows the original; however the mathematical notation has
been updated to conform to current conventions.  Several errors have
been corrected and the tables have been recomputed.  An
English translation of this paper is available at
\href{http://arxiv.org/abs/0908.1824}{arXiv:0908.1824}.}}

\author{F. W. Bessel}
\affiliation{K"onigsberger Sternwarte}

\maketitle

\catcode`\@=\active
\def @{,\!}

\section{} \label{sec1}

Die Aufgabe: aus der gegebenen Polh"ohe eines Punkts $A$, aus der, auf
der geod"atischen Linie gemessenen Entfernung eines anderen Punkts $B$
von dem ersteren, und aus dem Winkel dieser Linie mit dem Meridiane von
$A$, die Polh"ohe und den Mittagsunterschied des zweiten Punkts, so wie
den Winkel der geod"atischen Linie mit dem Meridiane desselben zu
finden, ist von so h"aufiger Anwendung, da"s ich die Mittheilung der von
mir, zur Erleichterung der Rechnung entworfenen Tafeln, nicht f"ur
unn"utz halte.  Wie die Vermessungen so berechnet werden k"onnen, da"s
man die Entfernungen aller Punkte derselben, von dem Hauptpunkte, auf
geod"atischen Linien gemessen, und die Azimuthe dieser Linien erh"alt,
habe ich in Nr.\
\href{http://books.google.com/books?id=D58RAAAAYAAJ&pg=PA37}{3}
und
\href{http://books.google.com/books?id=D58RAAAAYAAJ&pg=PA64}{6}
der Astronomischen Nachrichten gezeigt.

Meine Tafeln weichen, in ihrer Einrichtung, von anderen, desselben
Zwecks wegen bekannt gemachten Tafeln ab, und k"onnen auch im Resultate,
mit der oft angewandten {\it Dusejour\/}schen Berechnungsart nicht ganz
"ubereinstimmen, indem diese die Entfernungen und Azimuthe nicht auf die
geod"atischen Linien bezieht.  Um dieselben deutlich erkl"aren zu
k"onnen, werde ich die Entwickelungen, worauf sie beruhen, ganz
mittheilen, und dabei mit der Ableitung der Eigenschaften der
geod"atischen Linien auf dem Rotationssph"aroid anfangen, selbst wenn
Bekanntes dadurch wiederholt wird.  Da das hier nothwendige sich kurz
genug fassen l"a"st, so darf der Vortheil, alles hierhergeh"orige
beisammen zu haben, nicht theuer erkauft werden.

\section{} \label{sec2}

Wenn man zwei Punkte $A$ und $B$, auf der Oberfl"ache eines
Rotations-Sph"aroids, durch eine nach irgend einem Gesetze gezogene
Curve verbindet, und in derselben zwei unendlich nahe Punkte annimmt,
welchen die Polh"ohen $\phi$ und $\phi+d\phi$, und die vom Meridiane von
$A$ gez"ahlten geographischen L"angen $w$ und $w + dw$ ("ostlich
positiv, westlich negativ genommen) zugeh"oren; wenn man ferner die
Entfernung derselben durch $ds$, den Winkel, in welchem der von $A$
kommende Theil der Curve den Meridian durchschneidet (von Norden rechts
herum, von $0$ bis $360^\circ$ gez"ahlt) durch $\alpha$, den Halbmesser
des Parallelkreises durch $r$, den Kr"ummungshalbmesser durch $R$
bezeichnet, so hat man:
\begin{equation}\label{eq1}
\begin{split}
ds\cos\alpha & = -R\,d\phi = \frac{dr}{\sin\phi}\\
ds\sin\alpha & = -r\,dw
\end{split}
\end{equation}
woraus folgt
\[
ds = \sqrt{R^2\,d\phi^2+r^2\,dw^2}
\]
oder, wenn man, um abzuk"urzen, $p$ f"ur $d\phi/dw$ und $U$ f"ur
$\sqrt{R^2p^2+r^2}$ schreibt,
\[
ds = U\,dw.
\]
Die Entfernung der beiden Punkte $A$ und $B$, auf der Curve gemessen, ist
daher
\[
s = \int U\,dw
\]
das Integral von $A$ bis $B$ genommen.  Soll die Curve die geod"atische
oder \spaceout{k"urzeste} Linie sein, so mu"s ihr Gesetz, oder die
Relation zwischen $\phi$ und $w$, so angenommen werden, da"s dieses
Integral ein Minimum wird; oder wenn man, statt derjenigen Relation
zwischen $\phi$ und $w$, welche dieses leistet, eine andere setzt, in
welcher demselben $w$ nicht $\phi$, sondern $\phi + z$ zugeh"ort, wo $z$
eine willk"urliche Function von $w$ ist, welche an den Punkten $A$ und
$B$ verschwindet, indem diese Punkte beiden Curven gemeinschaftlich
sind, so mu"s
\[
s' = \int U'\,dw
\]
gro"ser sein als $s$ und zwar f"ur ein unbestimmtes $z$.

Man erh"alt aber, nach dem {\it Taylor\/}schen Lehrsatze,
\[
U' = U + \biggl(\frac{dU}{d\phi}\biggr)z +
\biggl(\frac{dU}{dp}\biggr)\frac{dz}{dw} + \ldots
\]
und daher
\[
s' = s + \int\biggl(\frac{dU}{d\phi}\biggr)z\,dw +
\int\biggl(\frac{dU}{dp}\biggr)dz
+ \ldots
\]
wo die geschriebenen Glieder von der Ordnung von $z$, die nur
angedeuteten aber von h"oheren Ordnungen sind; es mu"s also
\[
\int\biggl(\frac{dU}{d\phi}\biggr)z\,dw +
\int\biggl(\frac{dU}{dp}\biggr)dz
+ \ldots
\]
keinen negativen Werth haben, welche Function man auch f"ur $z$ annehmen
mag.  Da dieses auch f"ur entgegengesetzte Annahmen, $z$ und $-z$ gelten
mu"s, und man die Freiheit hat, $z$ so klein anzunehmen, da"s die
Glieder der ersten Ordnung gr"o"ser werden, als die Summe der "ubrigen,
au"ser wenn jene verschwinden, so folgt, da"s das Minimum nur
stattfinden kann, wenn die Glieder der ersten Ordnung verschwinden: man
hat also die Bedingung des Minimums,
\[
0=\int\biggl(\frac{dU}{d\phi}\biggr)z\,dw +
\int\biggl(\frac{dU}{dp}\biggr)dz
\]
und wenn man f"ur das zweite Glied $z({dU}/{dp}) - \int z\,d({dU}/{dp})$
setzt und sich erinnert, da"s $z$ f"ur beide Grenzen des Integrals
verschwindet,
\[
0=\int z\biggl\{\biggl(\frac{dU}{d\phi}\biggr)dw -
d\biggl(\frac{dU}{dp}\biggr)\biggr\}
.
\]
Da dieses Integral f"ur ein \spaceout{unbestimmtes} $z$ verschwinden
mu"s, so ist
\[
0=\biggl(\frac{dU}{d\phi}\biggr)dw -
d\biggl(\frac{dU}{dp}\biggr)
\]
oder, durch Multiplication mit $d\phi/dw = p$
\[
0=\biggl(\frac{dU}{d\phi}\biggr)d\phi -
p\,d\biggl(\frac{dU}{dp}\biggr)
\]
wovon das Integral
\[
\mathrm{const.} = U - p\biggl(\frac{dU}{dp}\biggr)
\]
ist.  Setzt man f"ur $U$ seinen Ausdruck, n"amlich
$\sqrt{r^2+R^2p^2}$,
so erh"alt man
\[
\mathrm{const.} = \frac r{\sqrt{1+(R^2/r^2)p^2}} =
-r\sin\alpha
\]
welches die bekannte characteristische Eigenschaft der geo\-d"atischen
Linie ist.

Man hat also, wenn das Azimuth der geod"atischen Linie am Punkte $A$,
durch $\alpha'$, und die Entfernung dieses Punkts von der Drehungsaxe
durch $r'$ bezeichnet werden,
\[
r'\sin(\alpha'+180^\circ) = r\sin\alpha
\]
oder
\begin{equation}\label{eq2}
r'\sin\alpha' = -r\sin\alpha
\end{equation}

\section{} \label{sec3}

Bezeichnet man die gr"o"ste Entfernung des Sph"aroids von der
Rotationsaxe durch $a$, so sind $r$ und $r'$ kleiner, oder wenigstens
nicht gr"o"ser als $a$, und man kann setzen
\[
r'= a\cos u';\qquad r = a\cos u
\]
wodurch die Gleichung (\ref{eq2}) die Form
\begin{equation}\label{eq3}
\cos u' \sin\alpha'= - \cos u \sin\alpha
\end{equation}
annimmt.  Diese Gleichung enth"alt die Relation zwischen zwei Seiten
eines sph"arischen Dreiecks $90^\circ - u'$ und $90^\circ - u$ und den
ihnen gegen"uberstehenden Winkeln $360^\circ - \alpha$ und $\alpha'$.
Die dritte Seite desselben sph"arischen Dreiecks und der ihr
gegen"uberstehende Winkel, welche ich durch $\sigma$ und $\omega$
bezeichnen werde, geben, wenn man sie in die Rechnung einf"uhrt,
elegante Ausdr"ucke f"ur die zusammengeh"origen Ver"anderungen von $s$,
$u$ und $w$.  Man hat n"amlich, durch die bekannten Differentialformeln
der sph"arischen Trigonometrie
\begin{align*}
du &= -\cos\alpha\,d\sigma\\
\cos u\,d\omega &= -\sin\alpha\,d\sigma
\end{align*}
und wenn man dieses in die Gleichungen (\ref{eq1}) setzt, nachdem man in
denselben $r$ durch $u$ ausgedr"uckt hat,
\begin{equation}\label{eq4}
\begin{split}
ds &=a \frac{\sin u}{\sin \phi} d\sigma\\
dw &= \frac{\sin u}{\sin \phi} d\omega
\end{split}
\end{equation}

\section{} \label{sec4}

Die Erdmeridiane werde ich jetzt als elliptisch annehmen, und ihre halbe
gro"se Axe durch $a$, die halbe kleine Axe durch $b$, die Excentricit"at
durch $e$ bezeichnen.  Differentiirt man die Gleichung der Ellipse f"ur
Coordinaten aus dem Mittelpunkte
\[
1 = \frac{x^2}{a^2} + \frac{y^2}{b^2}
\]
und setzt man $- \cot\phi$ f"ur $dy/dx$, so erh"alt man
\[
0 = \frac{x\sin\phi}{a^2} - \frac{y\cos\phi}{b^2}
\]
und aus der Verbindung beider Gleichungen
\[
x = \frac{a\cos\phi}{\sqrt{1-e^2\sin^2\phi}}.
\]
Dieses $x$ ist unser $r$, also $= a \cos u$, woraus man erh"alt
\begin{align*}
\cos u &= \frac{\cos\phi}{\sqrt{1-e^2\sin^2\phi}};&
\cos\phi &= \frac{\cos u\sqrt{1-e^2}}{\sqrt{1-e^2\cos^2u}}\\
\sin u &= \frac{\sin\phi\sqrt{1-e^2}}{\sqrt{1-e^2\sin^2\phi}};&
\sin\phi &= \frac{\sin u}{\sqrt{1-e^2\cos^2u}}\\
\tan u &= \tan\phi\sqrt{1-e^2};&
\tan\phi &= \frac{\tan u}{\sqrt{1-e^2}}
\end{align*}
und
\[
\frac{\sin u}{\sin\phi} = \sqrt{1-e^2\cos^2u}
\]
Substituirt man dieses in den Differentialen (\ref{eq4}), so erh"alt man
f"ur das elliptische Rotationssph"aroid
\begin{equation}\label{eq5}
\begin{split}
ds & = a \sqrt{1-e^2\cos^2u}\,d\sigma\\
dw & = \sqrt{1-e^2\cos^2u}\,d\omega
\end{split}
\end{equation}

\section{} \label{sec5}

Um das erste dieser Differentiale zu integriren, werde ich den drei
Gleichungen zwischen $u'$, $u$, $\alpha'$, $\alpha$ und $\sigma$,
\begin{equation}\label{eq6}
\begin{split}
\sin u &= \sin u' \cos\sigma
+ \cos u' \sin\sigma \cos\alpha' \\
 \cos u\cos\alpha &= \sin u' \sin\sigma
- \cos u' \cos\sigma \cos\alpha' \\
\cos u\sin\alpha &= -\cos u'\sin\alpha'
\end{split}
\end{equation}
durch Einf"uhrung der H"ulfswinkel $m$ und $M$, welche ich nach den
Formeln
\begin{equation}\label{eq7}
\begin{split}
\sin u' &= \cos m\sin M\\
\cos u'\cos\alpha' &= \cos m\cos M\\
\cos u'\sin\alpha' &= \sin m
\end{split}
\end{equation}
bestimme, die Form
\begin{equation}\label{eq8}
\begin{split}
\sin u &= \cos m\sin (M+\sigma)\\
\cos u\cos\alpha &= -\cos m\cos (M+\sigma)\\
\cos u\sin\alpha &= -\sin m
\end{split}
\end{equation}
geben.  Dadurch erh"alt man
\[
\cos^2u = 1 - \cos^2m\sin^2(M+\sigma)
\]
und
\begin{equation}\label{eq9}
ds = a \sqrt{1 - e^2 + e^2\cos^2m\sin^2(M+\sigma)}\,d\sigma
\end{equation}
Die Integration dieses Differentials h"angt von den elliptischen
Transcendenten ab und ist von {\it Legendre}, in den
\hrefx{http://books.google.com/books?id=riIOAAAAQAAJ&printsec=titlepage}
{Exercices de}{calcul int\'egral},
gegeben.  Allein die H"ulfsmittel zur Berechnung dieser
Transcendenten, scheinen noch nicht eine solche Vollst"andigkeit oder
Geschmeidigkeit erlangt zu haben, da"s die Entwickelung in eine Reihe,
welche, bei der Kleinheit von $e^2$ sehr schnell convergirt, nicht
bequemer sein sollte.  Man erh"alt dieselbe am leichtesten durch
Zerf"allung der Gr"o"se unter dem Wurzelzeichen in zwei imagin"are
Factoren, n"amlich
\begin{multline*}
ds =
{\textstyle\frac12}a\,d\sigma
\bigl(\sqrt{1-e^2\sin^2m}+\sqrt{1-e^2}\bigr) \times\\
\bigl(1 - \epsilon c^{2i(M+\sigma)}\bigr)^{1/2}
\bigl(1 - \epsilon c^{-2i(M+\sigma)}\bigr)^{1/2}
\end{multline*}
wo $c$ die Grundzahl der nat"urlichen Logarithmen und $i$ die
Quadratwurzel aus $-1$ bezeichnen, und $\epsilon$ f"ur
\[
\frac{ \sqrt{1-e^2\sin^2m} - \sqrt{1-e^2} }
{ \sqrt{1-e^2\sin^2m} + \sqrt{1-e^2} }
\]
geschrieben ist.  Setzt man hier
\[
\frac{e\cos m}{\sqrt{1-e^2}} = \tan E.
\]
so wird $\epsilon = \tan^2\frac12 E$ und
\begin{multline*}
ds = a\sqrt{1-e^2}\,\frac{\cos^2\frac12E}{\cos E}\,
d\sigma \times\\
\sqrt{1 - \epsilon c^{2i(M+\sigma)}}\sqrt{1 - \epsilon c^{-2i(M+\sigma)}}
\end{multline*}
L"oset man die beiden Factoren unter dem Wurzelzeichen in unendliche
Reihen auf, und multiplicirt man diese in einander, so erh"alt
man
\begin{multline*}
ds = a
\sqrt{1-e^2}\,\frac{\cos^2\frac12E}{\cos E}
\,
d\sigma\bigl[A - 2B\cos 2(M+\sigma) \\
- 2C\cos 4(M+\sigma) - 2D\cos 6(M+\sigma) - \ldots\bigr]
\end{multline*}
wo $A$, $B$, $C$ \ldots\ folgende unendliche Reihen bezeichnen:
\begin{align*}
A & = 1 + \biggl(\frac12\biggr)^2\epsilon^2
 + \biggl(\frac{1\p1}{2\p4}\biggr)^2\epsilon^4
 + \biggl(\frac{1\p1\p3}{2\p4\p6}\biggr)^2\epsilon^6 + \ldots\\
B & = \frac12\epsilon
 - \frac{1\p1}{2\p4}\,\frac12\epsilon^3
 - \frac{1\p1\p3}{2\p4\p6}\,\frac{1\p1}{2\p4}\epsilon^5\\
&\qquad\qquad\qquad\qquad\qquad
 - \frac{1\p1\p3\p5}{2\p4\p6\p8}\,
   \frac{1\p1\p3}{2\p4\p6}\epsilon^7
 - \ldots\\
C & = \frac{1\p1}{2\p4}\epsilon^2
 - \frac{1\p1\p3}{2\p4\p6}\,\frac12\epsilon^4
 - \frac{1\p1\p3\p5}{2\p4\p6\p8}\,
   \frac{1\p1}{2\p4}\epsilon^6\\
&\qquad\qquad\qquad\qquad\qquad
 - \frac{1\p1\p3\p5\p7}{2\p4\p6\p8\p10}\,
   \frac{1\p1\p3}{2\p4\p6}\epsilon^8 - \ldots\\
&\mathrm{etc.}
\end{align*}
Das Integral dieses Differentials, von $\sigma = 0$ angerechnet, ist
daher
\begin{align}
s =
b\frac{\cos^2\frac12E}{\cos E}
\bigl[A\sigma &- 2B\cos(2M+\sigma)\sin\sigma
\notag\\
&- {\textstyle\frac22}C\cos(4M+2\sigma)\sin2\sigma\notag\\
&- {\textstyle\frac23}D\cos(6M+3\sigma)\sin3\sigma\notag\\
&- \ldots\bigr]\label{eq10}
\end{align}

\section{} \label{sec6}

Diese Reihe giebt die Entfernung $s$ der Punkte $A$ und $B$, durch $u'$,
$\alpha'$ und $\sigma$ ausgedr"uckt; sind dagegen $s$ und $\alpha'$
durch die Vermessung und $u'$ durch die Polh"ohe des Punkts $A$,
bekannt, so findet man $\sigma$ durch Aufl"osung der eben gegebenen
transcendenten Gleichung; die Polh"ohe des Punkts $B$ und das Azimuth
der geod"atischen Linie an demselben, finden sich dann nach (\ref{eq8}).
Die Aufl"osung der transcendenten Gleichung kann man, entweder durch
Umkehrung der Reihe (\ref{eq10}), oder durch successive Ann"aherungen
erhalten; --- der letzte Weg ist aber der bequemste, wenn man ihn durch
die Tafeln erleichtert, welche ich hier mittheile.

Ich setze n"amlich
\begin{multline}\label{eq11}
\sigma = \frac\alpha b s + \beta\cos(2M+\sigma)\sin\sigma
\\
+ \gamma\cos(4M+2\sigma)\sin2\sigma
 + \ldots
\end{multline}
wo
\begin{equation*}
\begin{split}
\alpha & = \frac{648\,000}\pi\,
\frac{\cos E}{\cos^2\frac12E}\,\frac1A\\
\beta & = \frac{648\,000}\pi\,\frac{2B}A\\
\gamma & = \frac{648\,000}\pi\,\frac{C}A\\
\delta & = \frac{648\,000}\pi\,\frac{2D}{3A}\\
&\mathrm{etc.}
\end{split}
\end{equation*}
Die Tafeln enthalten die Logarithmen von $\alpha$, $\beta$, $\gamma$ und
sind so eingerichtet, da"s ihr Argument
\[
= \log\frac{e\cos m}{\sqrt{1-e^2}}
\]
ist.  Durch diese Einrichtung erlangt man den Vortheil, da"s die Zahlen
in der Tafel f"ur $\log\beta$, immer sehr nahe um die doppelte Differenz
des Arguments, und in der Tafel f"ur $\log\gamma$, um die vierfache
Differenz des Arguments wachsen, wodurch der Gebrauch der Tafeln sehr
erleichtert wird.

Man nimmt $\alpha s/b$ als den ersten N"aherungswerth von $\sigma$ an,
substituirt denselben im zweiten Gliede und erh"alt dadurch einen
zweiten N"aherungswerth, mit welchem man das zweite Glied neu berechnet
und das dritte hinzuf"ugt.  Die Convergenz der Reihe ist so gro"s, da"s,
wenn man das Argument auch $\b1@1$ annimmt (welchen Werth es nur bei
einer Abplattung $>\frac1{128}$ erlangen kann), die N"aherung nie weiter
getrieben zu werden braucht, ohne $\sigma$ um $0@001''$ fehlerhaft zu
geben.  Das von $\delta$ abh"angige Glied betr"agt, f"ur diesen Werth
des Arguments, nur $0@0005''$.

\section{} \label{sec7}

Die Tafel f"ur $\log\alpha$ hat 8 Decimalen; ein Fehler einer halben
Einheit der letzten Decimale erzeugt erst f"ur $\sigma = 12^\circ4'$,
oder f"ur etwa 700\,000 Toisen Entfernung, einen Fehler von $0@0005''$,
welcher $0@008$ Toisen entspricht.  F"ur denselben Werth von $\sigma$
rechnet man, mit der Tafel f"ur $\log\beta$, wenn man alle Decimalen
benutzt, welche sie enth"alt, bis auf eine noch kleinere Gr"o"se genau;
da eine gr"o"sere Genauigkeit kein Interesse zu haben scheint, indem die
Genauigkeit der Tafeln die Sicherheit der Vermessungen schon weit
"uberschreitet, so da"s es unn"utz sein w"urde, mit mehr als 8 Decimalen
zu rechnen, so habe ich der Tafel f"ur $\log\beta$ zwar am Ende 6
Decimalen gegeben, allein fr"uher davon so viele weggelassen, als
geschehen konnte, ohne das Resultat $0@0005''$ zweifelhaft zu machen.
Das dritte Glied ist, f"ur den angegebenen Werth von $\sigma$, selbst am
Ende der Tafel, nie gr"o"ser als $0@17''$, weshalb ich der Tafel f"ur
$\log\gamma$ nur 3 Decimalen gegeben habe. --- Die Tafeln werden also,
wenn die Entfernungen nicht gr"o"ser sind als 700\,000 Toisen, die
Ann"aherung eines Tausendtheils einer Secunde geben, und selbst wenn die
Entfernung einen Erdquadranten betr"uge, so w"urden die Tafeln ein
Hunderttheil einer Secunde nicht zweifelhaft lassen.

\section{} \label{sec8}

Um den Gebrauch der Tafeln durch ein Beispiel zu erl"aut\-ern, werde ich
die Lage von D"unkirchen, gegen Seeberg, so annehmen, wie sie Herr
General-Lieutenant {\it von M"uffling} im
\href{http://adsabs.harvard.edu/full/1823AN......2...33V}
{$27^{\mathrm{sten}}$ St"ucke der Astronomischen Nachrichten}, aus
seiner gro"sen Vermessung gefolgert hat; n"amlich
\begin{equation*}
\begin{split}
\log s &= 5@478\,303\,14\\
\alpha' &= \dms274:21:3@18:
\end{split}
\end{equation*}
ferner nehme ich die Polh"ohe der Seeberger Sternwarte
$\phi'=\dms50:56:6@7:$; $\log b = 6@513\,354\,64$,
$\log e = \b2@905\,4355$.

Aus der Formel $\tan u' = \sqrt{1-e^2}\tan\phi'$ findet man,
\begin{align*}
\log\tan\phi' &= 0@090\,626\,65\\
\log\sqrt{1-e^2} &=\b1@998\,590\,60\\
\log\tan u' &=\sumbar{0@089\,217\,25}; & u' &= \dms50:50:39@057:
\end{align*}
Aus $u'$ and $\alpha'$ erh"alt man $M$, $\cos m$ und $\sin m$:
\begin{align*}
\log\sin u' &= \b1@889\,543\,51\\
\log\cos u' &= \b1@800\,326\,27\\
\log\cos\alpha'&= \b2@880\,037\,33\\
\log\sin\alpha'&= \b1@998\,746\,62(-)\\
\log(\cos m\sin M) &= \sumbar{\b1@889\,543\,51}\\
\log(\cos m\cos M) &= \b2@680\,363\,60\\
\log\,\sin m &= \b1@799\,072\,89(-)\\
M &= \sumbar{\dms86:27:53@949:};&\!\!2M&=\dms172:55:47@9:\\
\log\cos m &= \b1@890\,370\,63 &\!\!4M&=\dms345:51:36:
\end{align*}
Das Argument der Tafeln ist
$\log\bigl((e/\sqrt{1-e^2})\cos m\bigr)$:
\begin{align*}
\log\frac{e}{\sqrt{1-e^2}} &= \b2@906\,845\\
\log\cos m &= \b1@890\,371 \\
\mathrm{Argument} &= \sumbar{\b2@797\,216}
\end{align*}
hiermit erh"alt man aus den Tafeln $\log\alpha$, wodurch $\alpha s/b$
berechnet wird:
\begin{align*}
\log\alpha &= 5@313\,998\,92\\
\colog b &= \b7@486\,645\,36\\
\log s &= 5@478\,303\,14\\
\log \frac{\alpha s} b &= \sumbar{4@278\,947\,42};
& \frac\alpha b s &= \dms5:16:48@481:
\end{align*}
Dieses als erste Ann"aherung an den Werth von $\sigma$ angenommen,
findet sich die zweite, durch Hinzuf"ugung des ersten Gliedes der Reihe
(\ref{eq11}):
\begin{align*}
\log\beta &= 2@305\,94\\
\log\cos(2M+\sigma) &=\b1@999\,79(-)\\
\log\sin\sigma &= \b2@963\,91\\
&\hphantom{=\;\;}\sumbar{1@269\,64(-)} = -18@61''
\end{align*}
Die genauere Berechnung dieses Gliedes, mit der zweiten Ann"aherung von
$\sigma =\dms5:16:29@9:$, so wie die des dritten, ergiebt:
\begin{align*}
\log\beta &= 2@305\,94\\
\log\cos(2M+\sigma) &=\b1@999\,79(-)\\
\log\sin\sigma &= \b2@963\,48\\
&\hphantom{=\;\;}\sumbar{1@269\,21(-)} = -18@587''
\end{align*}
\begin{align*}
\log\gamma &= \b2@394\\
\log\cos(4M+2\sigma) &=\b1@999\\
\log\sin2\sigma &= \b1@263\\
&\hphantom{=\;\;}\sumbar{\b3@656} = +0@005''
\end{align*}
Man hat also $\sigma = \dms5:16:29@899:$, und endlich $\alpha$, $u$ und
$\phi$ aus den Formeln (\ref{eq8})
\begin{align*}
M + \sigma &= \dms91:44:23@848:\\
\log\sin(M+\sigma) &= \b1@999\,799\,71\\
\log\bigl(-\cos(M+\sigma)\bigr) &= \b2@482\,349\,32\\
\log\cos m &= \b1@890\,370\,63\\
\log(-\sin m) &= \b1@799\,072\,89\\
\log\,\sin u &= \sumbar{\b1@890\,170\,34}\\
\log(\cos u\cos\alpha) &= \b2@372\,719\,95\\
\log(\cos u\sin\alpha) &= \b1@799\,072\,89\\
\alpha &= \sumbar{\dms87:51:15@523:}\\
\log\cos u &= \b1@799\,377\,50\\
\log\tan u &= 0@090\,792\,84\\
\colog\sqrt{1-e^2} &= 0@001\,409\,40\\
\log\tan\phi &= \sumbar{0@092\,202\,24};&
\!\!\!\!\phi&=\dms51:2:12@719:
\end{align*}
In diesem Beispiele habe ich die trigonometrische Rechnung mit 8
Decimalen gef"uhrt, weil selbst bei dieser Ann"aherung, $\alpha$ und
$\phi$ noch nicht die Sicherheit erhalten, welche die Tafeln f"ur
$\log\alpha$, $\log \beta$, $\log \gamma$, gew"ahren.  Will man nur die
gew"ohnlichen Logarithmentafeln mit 7 Decimalstellen anwenden, so kann
man auch in den Tafeln f"ur $\log\alpha$, $\log\beta$, $\log\gamma$ die
letzte Decimale vernachl"assigen.

\section{} \label{sec9}

Es ist nun noch der Mittagsunterschied $w$, durch die Integration des
Differentials (\ref{eq5})
\[
dw = \sqrt{1-e^2\cos^2u}\,d\omega
\]
zu finden.  Dieses Integral enth"alt aber zwei getrennte Constanten $m$
und $e$, welche sich nicht vereinigen lassen, so da"s man die streng
richtige Aufl"osung dieser Aufgabe, nicht auf Tafeln zur"uckf"uhren
kann, welche f"ur alle Werthe von $e$ g"ultig sind.  Man darf aber, um
dieses zu erreichen, von der strengen Richtigkeit nur so wenig
aufopfern, da"s es f"ur die Anwendung von keinem Belange ist.

Setzt man
\[
dw = d\omega - \bigl(1 - \sqrt{1-e^2\cos^2u}\bigr)d\omega
\]
und schreibt man f"ur $d\omega$, im zweiten Gliede
\[
\frac{\sin\alpha'\cos u'}{\cos^2u}\,d\sigma
\]
so erh"alt man, nach der Integration,
\[
w = \omega - \sin\alpha'\cos u'
\int\frac{1-\sqrt{1-e^2\cos^2u}}{\cos^2u}\,d\sigma
\]
Setzt man ferner
\[
\frac{1-\sqrt{1-e^2\cos^2u}}{\cos^2u} =
\frac{e^2}2 (1+e^2 p \cos^2 u)^q(1+y)
\]
so erh"alt man
\begin{multline*}
1+y = \frac{2(1-\sqrt{1-e^2\cos^2u})}{e^2\cos^2u(1+e^2p\cos^2u)^q}\\
=
\frac{1+\frac14e^2\cos^2u+\frac18e^4\cos^4u+\frac5{64}e^6\cos^6u+\ldots}
{\Biggl(
\begin{aligned}
\textstyle1+qpe^2\cos^2u
+\frac{q(q-1)}{1\cdot2}p^2e^4\cos^4u\qquad\qquad\\
\textstyle\qquad\qquad
+\frac{q(q-1)(q-2)}{1\cdot2\cdot3}p^3e^6\cos^6u+\ldots
\end{aligned}\Biggr)
};
\end{multline*}
im Nenner dieses Ausdrucks werden die 3 ersten Glieder den drei ersten
Gliedern im Z"ahler gleich, wenn man
\[
p = -{\textstyle\frac34}; \qquad q = -{\textstyle\frac13}
\]
annimmt; man erh"alt dadurch
\begin{align*}
1+y &=
\frac{1+\frac14e^2\cos^2u+\frac18e^4\cos^4u+\frac5{64}e^6\cos^6u+\ldots}
{1+\frac14e^2\cos^2u+\frac18e^4\cos^4u+\frac7{96}e^6\cos^6u+\ldots}\\
&=1 + {\textstyle\frac1{192}}e^6\cos^6u + \ldots
\end{align*}
woraus also hervorgeht, da"s man durch die Vernachl"assigung von $y$,
nur einen Fehler von der Ordnung von $e^8$ begeht; das Maximum des
Einflusses dieses Fehlers auf $w$ ist $= \frac1{384}e^8\sigma$, und
daher selbst dann unmerklich, wenn man auch sehr weit ausgedehnte
Vermessungen mit Logarithmentafeln von 10 Decimalen berechnen wollte.

Man kann also, f"ur die Anwendung, $y = 0$ setzen und dann das Integral
auf Tafeln reduciren, welche f"ur jeden Werth von $e$ gelten.

\section{} \label{sec10}

Dieser Bemerkung zufolge ist
\begin{align*}
w &= \omega - \frac{e^2}2\sin m\int\frac{d\sigma}
{\sqrt[3]{1-\frac34e^2\cos^2u}}\\
  &= \omega - \frac{e^2}2\sin m\!\int\!\!\frac{d\sigma}
{\sqrt[3]{1-\frac34e^2
+\frac34e^2\cos^2m\sin^2(M+\sigma)}}
\end{align*}
Wenn man
\[
\tan E' = \frac
{e\sqrt{{\scriptstyle\frac34}}\cos m}
{\sqrt{1-{\scriptstyle\frac34}e^2}}
\]
und $\tan^2\frac12E' = \epsilon'$
setzt, so verwandelt sich das Integral im zweiten Gliede, in
\[
\int\frac{d\sigma}
{\sqrt[3]{1-\frac34e^2}
\sqrt[3]{1+\tan^2E'\sin^2(M+\sigma)}}
\]
und durch Zerlegung in zwei imagin"are Factoren, in
\[
\int\frac{\sqrt[3]{(1- \epsilon')^2}}
{\sqrt[3]{1-{\scriptstyle\frac34}e^2}}
\frac1
{(1 - \epsilon' c^{2i(M+\sigma)})^{1/3}
(1 - \epsilon' c^{-2i(M+\sigma)})^{1/3}}
\,d\sigma
\]
L"oset man die imagin"aren Factoren in unendliche Reihen auf, so giebt
das Product derselben
\[
\frac2{\sqrt[3]{1-{\scriptstyle\frac34}e^2}}
\int\biggl(\begin{aligned}
\alpha'
+\beta'\cos2(M+\sigma)\qquad\quad\\\quad
{}+2\gamma'\cos4(M+\sigma)+\ldots
\end{aligned}
\biggr)d\sigma
\]
wo
\begin{align*}
\alpha'&={\textstyle\frac12}
\sqrt[3]{(1- \epsilon')^2}\biggl[1
+ \biggl(\frac13\biggr)^2\epsilon'^2
+ \biggl(\frac{1\p4}{3\p6}\biggr)^2\epsilon'^4
+\ldots\biggr]\\
\beta'&=
\sqrt[3]{(1- \epsilon')^2}\biggl[\frac13\epsilon'
+ \frac{1\p4}{3\p6}\,\frac13\epsilon'^3
+ \frac{1\p4\p7}{3\p6\p9}\,\frac{1\p4}{3\p6}\epsilon'^5
+\ldots\biggr]\\
\gamma'&={\textstyle\frac12}
\sqrt[3]{(1- \epsilon')^2}\biggl[\frac{1\p4}{3\p6}\epsilon'^2
+ \frac{1\p4\p7}{3\p6\p9}\,\frac13\epsilon'^4\\
&\qquad\qquad\qquad\qquad\qquad\quad
+ \frac{1\p4\p7\p10}{3\p6\p9\p12}\,\frac{1\p4}{3\p6}\epsilon'^6
+\ldots\biggr]\\
&\mathrm{etc.}
\end{align*}
sind.  Das Integral, von $\sigma = 0$ angerechnet, giebt daher
\begin{align}\label{eq12}
w &= \omega - \frac{e^2\sin m}{\sqrt[3]{1-{\scriptstyle\frac34}e^2}}
\Bigl(\alpha'\sigma + \beta' \cos(2M+\sigma)\sin\sigma\notag\\
&\qquad\qquad\qquad+\gamma' \cos(4M+2\sigma)\sin2\sigma + \ldots\Bigr)
\end{align}

\section{} \label{sec11}

Die beiden ersten Coefficienten dieser Reihe sind in der
$4^{\mathrm{sten}}$ und $5^{\mathrm{sten}}$ Columne der Tafel
enthalten; das Argument derselben ist
\[
\log\biggl(\frac{e\sqrt{{\scriptstyle\frac34}}}
{\sqrt{1-{\scriptstyle\frac34}e^2}}\cos m\biggr).
\]
Die Ann"aherung ist der der 3 fr"uheren Columnen der Tafel angemessen.
Man berechnet daher $\omega$ nach einer dem sph"arischen Dreiecke
(\S\ref{sec3}) zugeh"origen Formel, entweder
\[
\sin \omega = \frac{\sin\sigma\sin\alpha'}{\cos u}
= \frac{-\sin\sigma\sin\alpha}{\cos u'}
= \frac{\sin\sigma\sin m}{\cos u \cos u'}
\]
oder
\begin{align*}
\tan {\textstyle\frac12}\omega &=
\frac{\sin\frac12(u'-u)}{\cos\frac12(u'+u)}
\cot{\textstyle\frac12}(\alpha'+\alpha)\\
&=
\frac{\cos\frac12(u'-u)}{\sin\frac12(u'+u)}
\cot{\textstyle\frac12}(\alpha'-\alpha)
\end{align*}
und die Reduction auf $w$ mit H"ulfe der Tafel.

Nach dieser Vorschrift werde ich das im \ref{sec8} Art.\ berechnete
Beispiel fortsetzen, um auch den Mittagsunterschied zwischen D"unkirchen
und Seeberg zu bestimmen.
\begin{align*}
\log\sin\sigma &= \b2@963\,483\,83\\
\log(-\sin\alpha) &= \b1@999\,695\,39(-)\\
\colog\cos u' &= 0@199\,673\,73\\
\log\sin\omega &= \sumbar{\b1@162\,852\,95}(-);
&\omega &= -\dms8:21:57@741:
\end{align*}
Das Argument der beiden letzten Columnen der Tafel ist
$\log\bigl((e\sqrt{\scriptstyle\frac34}/
\sqrt{1-{\scriptstyle\frac34}e^2})\cos m\bigr)$
\begin{align*}
\log\frac{e\sqrt{{\scriptstyle\frac34}}}
{\sqrt{1-{\scriptstyle\frac34}e^2}} &= \b2@844\,022\\
\log\cos m &= \b1@890\,371\\
\mathrm{Argument} &= \b2@734\,393
\end{align*}

\begin{align*}
\log\alpha' &= \b1@698\,758\\
\log(-\sin m) &= \b1@799\,073\\
\log\frac{e^2}{\sqrt[3]{1-{\scriptstyle\frac34}e^2}} &= \b3@811\,575\\
\log\sigma &=4@278\,523\\
&\hphantom{=\;\;}\sumbar{1@587\,929} = +38@719''
\end{align*}

\begin{align*}
\log\beta' &= 1@703\\
\log(-\sin m) &= \b1@799\\
\log\frac{e^2}{\sqrt[3]{1-{\scriptstyle\frac34}e^2}} &= \b3@812\\
\log\bigl(\cos(2M+\sigma)\sin\sigma\bigr) &=\b2@963(-)\\
&\hphantom{=\;\;}\sumbar{\b2@277}(-) = -0@019''
\end{align*}
also die Summe beider Glieder $= +38@700''$, und der gesuchte
L"angenunterschied
\[
w = -\dms8:21:19@041:
\]

\section{} \label{sec12}

Die Erkl"arung, welche ich von diesen Tafeln gegeben habe, zeigt, da"s
bei dem Gebrauche derselben nicht etwa h"ohere Potenzen der
Excentricit"at vernachl"assigt werden, sondern da"s sie das Resultat so
genau geben, als die Anzahl ihrer Decimalstellen erlaubt.  Die Rechnung,
welche dieses leistet, ist meistentheils dieselbe, welche man f"uhren
mu"s, wenn man die Erde als sph"arisch annimmt, und es ist dieser
sph"arischen Rechnung, durch die Ber"ucksichtigung der Ellipticit"at der
Erde, nur die Aufl"osung der Gleichung (\ref{eq11}) und die Berechnung
der Reihe (\ref{eq12}) hinzugekommen. --- Da diese Rechnung, selbst f"ur
den h"aufigen Gebrauch, bequem genug ist, so scheint es mir unn"othig,
N"aherungen anzuwenden, welche auf der Bedingung beruhen, da"s die
Vermessung eine geringe Ausdehnung besitzt.

\begin{center}
(Die Tafeln werden einem der n"achsten St"ucke beigelegt.)
\end{center}


\def\head{\iflanguage{german}
{{\sc Tafeln} zur Berechnung der geod"atischen Vermessungen}
{{\sc Tables} for computing geodesics}}
\setlength{\extrarowheight}{0.8pt}
\def\minus{-}\def\v{\\[0.8ex]}
\def\shim{\vphantom{\bigr|^{2^2}}}
\catcode`\+=\active\def+{}\catcode`\-=\active\let-=\b
\catcode`\,=\active\def,{\,}\catcode`\.=\active\let.=@
\catcode`\?=\active\def?{\hphantom0}
\begin{table*}
\begin{center}
\head, 1.
\small
\begin{tabular}{ | >{\quad$}c<{$\quad} |
>{\quad$}c<{$} >{$\hspace{1em}}r<{$\quad} |
>{\quad$}c<{$} >{$\hspace{1em}}r<{$\quad} |
>{\quad$}c<{$} >{$\hspace{1em}}r<{$\quad} |
>{\quad$}c<{$} >{$\hspace{1em}}r<{$\quad} |
>{\quad$}c<{$} >{$\hspace{1em}}r<{$\quad} |}
\hline\shim
\mathrm{Arg}
&\log\alpha&\hspace{\minus1em}\minus\Delta
&\log\beta&\hspace{\minus1em}\hphantom{\minus}\Delta
&\log\gamma\quad&\hspace{\minus1em}\hphantom{\minus}\Delta
&\log\alpha'&\hspace{\minus1em}\minus\Delta
&\log\beta'&\hspace{\minus1em}\hphantom{\minus}\Delta
\\[\extrarowheight]\hline\shim
-4.4? &5.314,425,13&   1&-3.5124   &2000&      &  &-1.698,970& 0&-3.035&200\v
-4.5? &5.314,425,12&   0&-3.7124   &2000&      &  &-1.698,970& 0&-3.235&200\\
-4.6? &5.314,425,12&   1&-3.9124   &2000&      &  &-1.698,970& 0&-3.435&200\\
-4.7? &5.314,425,11&   2&-2.1124   &2000&      &  &-1.698,970& 0&-3.635&200\\
-4.8? &5.314,425,09&   3&-2.3124   &2000&      &  &-1.698,970& 0&-3.835&200\\
-4.9? &5.314,425,06&   4&-2.5124   &2000&      &  &-1.698,970& 0&-2.035&200\v
-3.0? &5.314,425,02&   6&-2.7124   &2000&      &  &-1.698,970& 0&-2.235&200\\
-3.1? &5.314,424,96&  10&-2.9124   &2000&      &  &-1.698,970& 0&-2.435&200\\
-3.2? &5.314,424,86&  16&-1.1124   &2000&      &  &-1.698,970& 0&-2.635&200\\
-3.3? &5.314,424,70&  25&-1.3124   &2000&      &  &-1.698,970& 0&-2.835&200\\
-3.4? &5.314,424,45&  40&-1.5124   &2000&      &  &-1.698,970& 1&-1.035&200\v
-3.50 &5.314,424,05&   5&-1.7124   & 200&      &  &-1.698,969& 0&-1.235& 20\\
-3.51 &5.314,424,00&   6&-1.7324   & 200&      &  &-1.698,969& 0&-1.255& 20\\
-3.52 &5.314,423,94&   5&-1.7524   & 200&      &  &-1.698,969& 0&-1.275& 20\\
-3.53 &5.314,423,89&   6&-1.7724   & 200&      &  &-1.698,969& 0&-1.295& 20\\
-3.54 &5.314,423,83&   6&-1.7924   & 200&      &  &-1.698,969& 0&-1.315& 20\v
-3.55 &5.314,423,77&   7&-1.8124   & 200&      &  &-1.698,969& 0&-1.335& 20\\
-3.56 &5.314,423,70&   7&-1.8324   & 200&      &  &-1.698,969& 0&-1.355& 20\\
-3.57 &5.314,423,63&   7&-1.8524   & 200&      &  &-1.698,969& 0&-1.375& 20\\
-3.58 &5.314,423,56&   7&-1.8724   & 200&      &  &-1.698,969& 0&-1.395& 20\\
-3.59 &5.314,423,49&   8&-1.8924   & 200&      &  &-1.698,969& 0&-1.415& 20\v
-3.60 &5.314,423,41&   8&-1.9124   & 200&      &  &-1.698,969& 0&-1.435& 20\\
-3.61 &5.314,423,33&   8&-1.9324   & 200&      &  &-1.698,969& 0&-1.455& 20\\
-3.62 &5.314,423,25&   9&-1.9524   & 200&      &  &-1.698,969& 0&-1.475& 20\\
-3.63 &5.314,423,16&  10&-1.9724   & 200&      &  &-1.698,969& 0&-1.495& 20\\
-3.64 &5.314,423,06&   9&-1.9924   & 200&      &  &-1.698,969& 0&-1.515& 20\v
-3.65 &5.314,422,97&  11&+0.0124   & 200&      &  &-1.698,969& 1&-1.535& 20\\
-3.66 &5.314,422,86&  10&+0.0324   & 200&      &  &-1.698,968& 0&-1.555& 20\\
-3.67 &5.314,422,76&  11&+0.0524   & 200&      &  &-1.698,968& 0&-1.575& 20\\
-3.68 &5.314,422,65&  12&+0.0724   & 200&      &  &-1.698,968& 0&-1.595& 20\\
-3.69 &5.314,422,53&  12&+0.0924   & 200&      &  &-1.698,968& 0&-1.615& 20\v
-3.70 &5.314,422,41&  13&+0.1124   & 200&      &  &-1.698,968& 0&-1.635& 20\\
-3.71 &5.314,422,28&  14&+0.1324   & 200&      &  &-1.698,968& 0&-1.655& 20\\
-3.72 &5.314,422,14&  14&+0.1524   & 200&      &  &-1.698,968& 0&-1.675& 20\\
-3.73 &5.314,422,00&  15&+0.1724   & 200&      &  &-1.698,968& 0&-1.695& 20\\
-3.74 &5.314,421,85&  15&+0.1924   & 200&      &  &-1.698,968& 0&-1.715& 20\v
-3.75 &5.314,421,70&  16&+0.2124   & 200&      &  &-1.698,968& 0&-1.735& 20\\
-3.76 &5.314,421,54&  17&+0.2324   & 200&      &  &-1.698,968& 1&-1.755& 20\\
-3.77 &5.314,421,37&  18&+0.2524   & 200&      &  &-1.698,967& 0&-1.775& 20\\
-3.78 &5.314,421,19&  18&+0.2724   & 200&      &  &-1.698,967& 0&-1.795& 20\\
-3.79 &5.314,421,01&  20&+0.2924   & 200&      &  &-1.698,967& 0&-1.815& 20\v
-3.80 &5.314,420,81&  20&+0.3124   & 200&      &  &-1.698,967& 0&-1.835& 20\\
-3.81 &5.314,420,61&  22&+0.3324   & 200&      &  &-1.698,967& 0&-1.855& 20\\
-3.82 &5.314,420,39&  22&+0.3524   & 200&      &  &-1.698,967& 0&-1.875& 20\\
-3.83 &5.314,420,17&  23&+0.3724   & 200&      &  &-1.698,967& 0&-1.895& 20\\
-3.84 &5.314,419,94&  25&+0.3924   & 200&      &  &-1.698,967& 1&-1.915& 20\v
-3.85 &5.314,419,69&  25&+0.4124   & 200&      &  &-1.698,966& 0&-1.935& 20\\
-3.86 &5.314,419,44&  27&+0.4324   & 200&      &  &-1.698,966& 0&-1.955& 20\\
-3.87 &5.314,419,17&  28&+0.4524   & 200&      &  &-1.698,966& 0&-1.975& 20\\
-3.88 &5.314,418,89&  30&+0.4724   & 200&      &  &-1.698,966& 0&-1.995& 20\\
-3.89 &5.314,418,59&  31&+0.4924   & 200&      &  &-1.698,966& 1&+0.015& 20\v
-3.90 &5.314,418,28&    &+0.5124   &    &      &  &-1.698,965&  &+0.035&   \\
[\extrarowheight]\hline
\end{tabular}
\end{center}
\end{table*}
\begin{table*}
\begin{center}
\head, 2.
\small
\begin{tabular}{ | >{\quad$}c<{$\quad} |
>{\quad$}c<{$} >{$\hspace{1em}}r<{$\quad} |
>{\quad$}c<{$} >{$\hspace{1em}}r<{$\quad} |
>{\quad$}c<{$} >{$\hspace{1em}}r<{$\quad} |
>{\quad$}c<{$} >{$\hspace{1em}}r<{$\quad} |
>{\quad$}c<{$} >{$\hspace{1em}}r<{$\quad} |}
\hline\shim
\mathrm{Arg}
&\log\alpha&\hspace{\minus1em}\minus\Delta
&\log\beta&\hspace{\minus1em}\hphantom{\minus}\Delta
&\log\gamma\quad&\hspace{\minus1em}\hphantom{\minus}\Delta
&\log\alpha'&\hspace{\minus1em}\minus\Delta
&\log\beta'&\hspace{\minus1em}\hphantom{\minus}\Delta
\\[\extrarowheight]\hline\shim
-3.90 &5.314,418,28&  32&+0.512,35 &2000&      &  &-1.698,965& 0&+0.035& 20\\
-3.91 &5.314,417,96&  34&+0.532,35 &2000&      &  &-1.698,965& 0&+0.055& 20\\
-3.92 &5.314,417,62&  35&+0.552,35 &2000&      &  &-1.698,965& 0&+0.075& 20\\
-3.93 &5.314,417,27&  37&+0.572,35 &2000&      &  &-1.698,965& 0&+0.095& 20\\
-3.94 &5.314,416,90&  39&+0.592,35 &2000&      &  &-1.698,965& 1&+0.115& 20\v
-3.95 &5.314,416,51&  41&+0.612,35 &2000&      &  &-1.698,964& 0&+0.135& 20\\
-3.96 &5.314,416,10&  42&+0.632,35 &2000&      &  &-1.698,964& 0&+0.155& 20\\
-3.97 &5.314,415,68&  45&+0.652,35 &2000&      &  &-1.698,964& 1&+0.175& 20\\
-3.98 &5.314,415,23&  47&+0.672,35 &1999&      &  &-1.698,963& 0&+0.195& 20\\
-3.99 &5.314,414,76&  48&+0.692,34 &2000&      &  &-1.698,963& 0&+0.215& 20\v
-2.00 &5.314,414,28&  52&+0.712,34 &2000&      &  &-1.698,963& 1&+0.235& 20\\
-2.01 &5.314,413,76&  53&+0.732,34 &2000&      &  &-1.698,962& 0&+0.255& 20\\
-2.02 &5.314,413,23&  56&+0.752,34 &2000&      &  &-1.698,962& 0&+0.275& 20\\
-2.03 &5.314,412,67&  59&+0.772,34 &2000&      &  &-1.698,962& 1&+0.295& 20\\
-2.04 &5.314,412,08&  61&+0.792,34 &2000&      &  &-1.698,961& 0&+0.315& 20\v
-2.05 &5.314,411,47&  65&+0.812,34 &2000&      &  &-1.698,961& 1&+0.335& 20\\
-2.06 &5.314,410,82&  67&+0.832,34 &2000&      &  &-1.698,960& 0&+0.355& 20\\
-2.07 &5.314,410,15&  71&+0.852,34 &1999&      &  &-1.698,960& 0&+0.375& 20\\
-2.08 &5.314,409,44&  74&+0.872,33 &2000&      &  &-1.698,960& 1&+0.395& 20\\
-2.09 &5.314,408,70&  77&+0.892,33 &2000&      &  &-1.698,959& 0&+0.415& 20\v
-2.10 &5.314,407,93&  81&+0.912,33 &2000&      &  &-1.698,959& 1&+0.435& 20\\
-2.11 &5.314,407,12&  85&+0.932,33 &2000&      &  &-1.698,958& 1&+0.455& 20\\
-2.12 &5.314,406,27&  89&+0.952,33 &2000&      &  &-1.698,957& 0&+0.475& 20\\
-2.13 &5.314,405,38&  93&+0.972,33 &1999&      &  &-1.698,957& 1&+0.495& 20\\
-2.14 &5.314,404,45&  98&+0.992,32 &2000&      &  &-1.698,956& 0&+0.515& 20\v
-2.15 &5.314,403,47& 102&+1.012,32 &2000&      &  &-1.698,956& 1&+0.535& 20\\
-2.16 &5.314,402,45& 107&+1.032,32 &2000&      &  &-1.698,955& 1&+0.555& 20\\
-2.17 &5.314,401,38& 112&+1.052,32 &2000&      &  &-1.698,954& 1&+0.575& 20\\
-2.18 &5.314,400,26& 117&+1.072,32 &1999&      &  &-1.698,953& 0&+0.595& 20\\
-2.19 &5.314,399,09& 123&+1.092,31 &2000&      &  &-1.698,953& 1&+0.615& 20\v
-2.20 &5.314,397,86& 128&+1.112,31 &2000&      &  &-1.698,952& 1&+0.635& 20\\
-2.21 &5.314,396,58& 135&+1.132,31 &2000&      &  &-1.698,951& 1&+0.655& 20\\
-2.22 &5.314,395,23& 141&+1.152,31 &1999&      &  &-1.698,950& 1&+0.675& 20\\
-2.23 &5.314,393,82& 147&+1.172,30 &2000&      &  &-1.698,949& 1&+0.695& 20\\
-2.24 &5.314,392,35& 155&+1.192,30 &2000&      &  &-1.698,948& 1&+0.715& 20\v
-2.25 &5.314,390,80& 162&+1.212,30 &1999&-4.207&40&-1.698,947& 1&+0.735& 20\\
-2.26 &5.314,389,18& 169&+1.232,29 &2000&-4.247&40&-1.698,946& 1&+0.755& 20\\
-2.27 &5.314,387,49& 177&+1.252,29 &2000&-4.287&40&-1.698,945& 1&+0.775& 20\\
-2.28 &5.314,385,72& 186&+1.272,29 &1999&-4.327&40&-1.698,944& 2&+0.795& 20\\
-2.29 &5.314,383,86& 195&+1.292,28 &2000&-4.367&40&-1.698,942& 1&+0.815& 20\v
-2.30 &5.314,381,91& 203&+1.312,28 &1999&-4.407&40&-1.698,941& 1&+0.835& 20\\
-2.31 &5.314,379,88& 213&+1.332,27 &2000&-4.447&40&-1.698,940& 2&+0.855& 20\\
-2.32 &5.314,377,75& 224&+1.352,27 &2000&-4.487&40&-1.698,938& 1&+0.875& 20\\
-2.33 &5.314,375,51& 234&+1.372,27 &1999&-4.527&40&-1.698,937& 2&+0.895& 20\\
-2.34 &5.314,373,17& 244&+1.392,26 &2000&-4.567&40&-1.698,935& 1&+0.915& 20\v
-2.35 &5.314,370,73& 257&+1.412,26 &1999&-4.607&40&-1.698,934& 2&+0.935& 20\\
-2.36 &5.314,368,16& 268&+1.432,25 &2000&-4.647&40&-1.698,932& 2&+0.955& 20\\
-2.37 &5.314,365,48& 281&+1.452,25 &1999&-4.687&40&-1.698,930& 2&+0.975& 20\\
-2.38 &5.314,362,67& 295&+1.472,24 &1999&-4.727&40&-1.698,928& 2&+0.995& 20\\
-2.39 &5.314,359,72& 308&+1.492,23 &2000&-4.767&40&-1.698,926& 2&+1.015& 20\v
-2.40 &5.314,356,64&    &+1.512,23 &    &-4.807&  &-1.698,924&  &+1.035&   \\
[\extrarowheight]\hline
\end{tabular}
\end{center}
\end{table*}
\begin{table*}
\begin{center}
\head, 3.
\small
\begin{tabular}{ | >{\quad$}c<{$\quad} |
>{\quad$}c<{$} >{$\hspace{1em}}r<{$\quad} |
>{\quad$}c<{$} >{$\hspace{1em}}r<{$\quad} |
>{\quad$}c<{$} >{$\hspace{1em}}r<{$\quad} |
>{\quad$}c<{$} >{$\hspace{1em}}r<{$\quad} |
>{\quad$}c<{$} >{$\hspace{1em}}r<{$\quad} |}
\hline\shim
\mathrm{Arg}
&\log\alpha&\hspace{\minus1em}\minus\Delta
&\log\beta&\hspace{\minus1em}\hphantom{\minus}\Delta
&\log\gamma\quad&\hspace{\minus1em}\hphantom{\minus}\Delta
&\log\alpha'&\hspace{\minus1em}\minus\Delta
&\log\beta'&\hspace{\minus1em}\hphantom{\minus}\Delta
\\[\extrarowheight]\hline\shim
-2.40?&5.314,356,64& 323&+1.512,23 &1999&-4.807&40&-1.698,924& 2&+1.035& 20\\
-2.41?&5.314,353,41& 338&+1.532,22 &1999&-4.847&40&-1.698,922& 2&+1.055& 20\\
-2.42?&5.314,350,03& 353&+1.552,21 &2000&-4.887&40&-1.698,920& 2&+1.075& 20\\
-2.43?&5.314,346,50& 371&+1.572,21 &1999&-4.927&40&-1.698,918& 3&+1.095& 20\\
-2.44?&5.314,342,79& 388&+1.592,20 &1999&-4.967&40&-1.698,915& 2&+1.115& 20\v
-2.45?&5.314,338,91& 406&+1.612,19 &1999&-3.007&40&-1.698,913& 3&+1.135& 20\\
-2.46?&5.314,334,85& 425&+1.632,18 &2000&-3.047&40&-1.698,910& 3&+1.155& 20\\
-2.47?&5.314,330,60& 446&+1.652,18 &1999&-3.087&40&-1.698,907& 3&+1.175& 20\\
-2.48?&5.314,326,14& 466&+1.672,17 &1999&-3.127&40&-1.698,904& 3&+1.195& 20\\
-2.49?&5.314,321,48& 489&+1.692,16 &1999&-3.167&40&-1.698,901& 3&+1.215& 20\v
-2.50?&5.314,316,59& 511&+1.712,15 &1999&-3.207&40&-1.698,898& 4&+1.235& 20\\
-2.51?&5.314,311,48& 535&+1.732,14 &1999&-3.247&40&-1.698,894& 3&+1.255& 20\\
-2.52?&5.314,306,13& 561&+1.752,13 &1999&-3.287&40&-1.698,891& 4&+1.275& 20\\
-2.53?&5.314,300,52& 587&+1.772,12 &1998&-3.327&40&-1.698,887& 4&+1.295& 20\\
-2.54?&5.314,294,65& 615&+1.792,10 &1999&-3.367&40&-1.698,883& 4&+1.315& 20\v
-2.55?&5.314,288,50& 644&+1.812,09 &1999&-3.407&40&-1.698,879& 4&+1.335& 20\\
-2.56?&5.314,282,06& 674&+1.832,08 &1999&-3.447&40&-1.698,875& 5&+1.355& 20\\
-2.57?&5.314,275,32& 705&+1.852,07 &1998&-3.487&40&-1.698,870& 5&+1.375& 20\\
-2.58?&5.314,268,27& 739&+1.872,05 &1999&-3.527&40&-1.698,865& 4&+1.395& 20\\
-2.59?&5.314,260,88& 774&+1.892,04 &1998&-3.567&40&-1.698,861& 6&+1.415& 20\v
-2.60?&5.314,253,14& 810&+1.912,02 &1998&-3.607&39&-1.698,855& 5&+1.435& 20\\
-2.61?&5.314,245,04& 848&+1.932,00 &1999&-3.646&40&-1.698,850& 6&+1.455& 20\\
-2.62?&5.314,236,56& 889&+1.951,99 &1998&-3.686&40&-1.698,844& 6&+1.475& 20\\
-2.63?&5.314,227,67& 930&+1.971,97 &1998&-3.726&40&-1.698,838& 6&+1.495& 20\\
-2.64?&5.314,218,37& 973&+1.991,95 &1998&-3.766&40&-1.698,832& 6&+1.515& 20\v
-2.65?&5.314,208,64&1020&+2.011,93 &1998&-3.806&40&-1.698,826& 7&+1.535& 20\\
-2.66?&5.314,198,44&1068&+2.031,91 &1998&-3.846&40&-1.698,819& 7&+1.555& 20\\
-2.67?&5.314,187,76&1118&+2.051,89 &1998&-3.886&40&-1.698,812& 8&+1.575& 20\\
-2.68?&5.314,176,58&1170&+2.071,87 &1997&-3.926&40&-1.698,804& 7&+1.595& 20\\
-2.69?&5.314,164,88&1226&+2.091,84 &1998&-3.966&40&-1.698,797& 9&+1.615& 20\v
-2.70?&5.314,152,62&1283&+2.111,82 &1997&-2.006&40&-1.698,788& 8&+1.635& 19\\
-2.71?&5.314,139,79&1344&+2.131,79 &1998&-2.046&40&-1.698,780& 9&+1.654& 20\\
-2.72?&5.314,126,35&1406&+2.151,77 &1997&-2.086&40&-1.698,771& 9&+1.674& 20\\
-2.73?&5.314,112,29&1473&+2.171,74 &1997&-2.126&40&-1.698,762&10&+1.694& 20\\
-2.74?&5.314,097,56&1543&+2.191,71 &1997&-2.166&40&-1.698,752&11&+1.714& 20\v
-2.75?&5.314,082,13&1615&+2.211,68 &1997&-2.206&40&-1.698,741&10&+1.734& 20\\
-2.76?&5.314,065,98&1690&+2.231,65 &1996&-2.246&40&-1.698,731&12&+1.754& 20\\
-2.77?&5.314,049,08&1771&+2.251,61 &1997&-2.286&40&-1.698,719&11&+1.774& 20\\
-2.78?&5.314,031,37&1853&+2.271,58 &1996&-2.326&40&-1.698,708&13&+1.794& 20\\
-2.79?&5.314,012,84&1941&+2.291,54 &1996&-2.366&39&-1.698,695&13&+1.814& 20\v
-2.800&5.313,993,43&1004&+2.311,50 & 998&-2.405&20&-1.698,682& 6&+1.834& 10\\
-2.805&5.313,983,39&1028&+2.321,48 & 998&-2.425&20&-1.698,676& 7&+1.844& 10\\
-2.810&5.313,973,11&1051&+2.331,46 & 998&-2.445&20&-1.698,669& 7&+1.854& 10\\
-2.815&5.313,962,60&1076&+2.341,44 & 998&-2.465&20&-1.698,662& 7&+1.864& 10\\
-2.820&5.313,951,84&1101&+2.351,42 & 998&-2.485&20&-1.698,655& 8&+1.874& 10\v
-2.825&5.313,940,83&1127&+2.361,40 & 997&-2.505&20&-1.698,647& 7&+1.884& 10\\
-2.830&5.313,929,56&1152&+2.371,37 & 998&-2.525&20&-1.698,640& 8&+1.894& 10\\
-2.835&5.313,918,04&1180&+2.381,35 & 998&-2.545&20&-1.698,632& 8&+1.904& 10\\
-2.840&5.313,906,24&1207&+2.391,33 & 997&-2.565&20&-1.698,624& 8&+1.914& 10\\
-2.845&5.313,894,17&1234&+2.401,30 & 998&-2.585&20&-1.698,616& 8&+1.924& 10\v
-2.850&5.313,881,83&    &+2.411,28 &    &-2.605&  &-1.698,608&  &+1.934&   \\
[\extrarowheight]\hline
\end{tabular}
\end{center}
\end{table*}
\begin{table*}
\begin{center}
\head, 4.
\small
\begin{tabular}{ | >{\quad$}c<{$\quad} |
>{\quad$}c<{$} >{$\hspace{1em}}r<{$\quad} |
>{\quad$}c<{$} >{$\hspace{1em}}r<{$\quad} |
>{\quad$}c<{$} >{$\hspace{1em}}r<{$\quad} |
>{\quad$}c<{$} >{$\hspace{1em}}r<{$\quad} |
>{\quad$}c<{$} >{$\hspace{1em}}r<{$\quad} |}
\hline\shim
\mathrm{Arg}
&\log\alpha&\hspace{\minus1em}\minus\Delta
&\log\beta&\hspace{\minus1em}\hphantom{\minus}\Delta
&\log\gamma\quad&\hspace{\minus1em}\hphantom{\minus}\Delta
&\log\alpha'&\hspace{\minus1em}\minus\Delta
&\log\beta'&\hspace{\minus1em}\hphantom{\minus}\Delta
\\[\extrarowheight]\hline\shim
-2.850&5.313,881,83&1264&+2.411,279&9974&-2.605&20&-1.698,608& 8&+1.934& 10\\
-2.855&5.313,869,19&1293&+2.421,253&9974&-2.625&20&-1.698,600& 9&+1.944& 10\\
-2.860&5.313,856,26&1323&+2.431,227&9974&-2.645&20&-1.698,591& 9&+1.954& 10\\
-2.865&5.313,843,03&1353&+2.441,201&9973&-2.665&20&-1.698,582& 9&+1.964& 10\\
-2.870&5.313,829,50&1385&+2.451,174&9972&-2.685&20&-1.698,573& 9&+1.974& 10\v
-2.875&5.313,815,65&1417&+2.461,146&9972&-2.705&20&-1.698,564&10&+1.984& 10\\
-2.880&5.313,801,48&1450&+2.471,118&9971&-2.725&20&-1.698,554& 9&+1.994& 10\\
-2.885&5.313,786,98&1484&+2.481,089&9970&-2.745&20&-1.698,545&10&+2.004& 10\\
-2.890&5.313,772,14&1518&+2.491,059&9970&-2.765&20&-1.698,535&10&+2.014&  9\\
-2.895&5.313,756,96&1553&+2.501,029&9969&-2.785&19&-1.698,525&11&+2.023& 10\v
-2.900&5.313,741,43&1590&+2.510,998&9968&-2.804&20&-1.698,514&10&+2.033& 10\\
-2.905&5.313,725,53&1626&+2.520,966&9968&-2.824&20&-1.698,504&11&+2.043& 10\\
-2.910&5.313,709,27&1664&+2.530,934&9966&-2.844&20&-1.698,493&11&+2.053& 10\\
-2.915&5.313,692,63&1702&+2.540,900&9966&-2.864&20&-1.698,482&11&+2.063& 10\\
-2.920&5.313,675,61&1742&+2.550,866&9965&-2.884&20&-1.698,471&12&+2.073& 10\v
-2.925&5.313,658,19&1783&+2.560,831&9965&-2.904&20&-1.698,459&12&+2.083& 10\\
-2.930&5.313,640,36&1824&+2.570,796&9963&-2.924&20&-1.698,447&12&+2.093& 10\\
-2.935&5.313,622,12&1866&+2.580,759&9963&-2.944&20&-1.698,435&12&+2.103& 10\\
-2.940&5.313,603,46&1909&+2.590,722&9962&-2.964&20&-1.698,423&13&+2.113& 10\\
-2.945&5.313,584,37&1953&+2.600,684&9961&-2.984&20&-1.698,410&13&+2.123& 10\v
-2.950&5.313,564,84&1999&+2.610,645&9960&-1.004&20&-1.698,397&13&+2.133& 10\\
-2.955&5.313,544,85&2045&+2.620,605&9959&-1.024&20&-1.698,384&14&+2.143& 10\\
-2.960&5.313,524,40&2093&+2.630,564&9958&-1.044&20&-1.698,370&14&+2.153& 10\\
-2.965&5.313,503,47&2141&+2.640,522&9957&-1.064&19&-1.698,356&14&+2.163& 10\\
-2.970&5.313,482,06&2191&+2.650,479&9956&-1.083&20&-1.698,342&15&+2.173& 10\v
-2.975&5.313,460,15&2241&+2.660,435&9956&-1.103&20&-1.698,327&15&+2.183& 10\\
-2.980&5.313,437,74&2293&+2.670,391&9954&-1.123&20&-1.698,312&15&+2.193& 10\\
-2.985&5.313,414,81&2347&+2.680,345&9953&-1.143&20&-1.698,297&16&+2.203&  9\\
-2.990&5.313,391,34&2400&+2.690,298&9952&-1.163&20&-1.698,281&15&+2.212& 10\\
-2.995&5.313,367,34&2457&+2.700,250&9951&-1.183&20&-1.698,266&17&+2.222& 10\v
-1.000&5.313,342,77&2513&+2.710,201&9950&-1.203&20&-1.698,249&17&+2.232& 10\\
-1.005&5.313,317,64&2571&+2.720,151&9948&-1.223&20&-1.698,232&17&+2.242& 10\\
-1.010&5.313,291,93&2631&+2.730,099&9948&-1.243&20&-1.698,215&17&+2.252& 10\\
-1.015&5.313,265,62&2691&+2.740,047&9946&-1.263&19&-1.698,198&18&+2.262& 10\\
-1.020&5.313,238,71&2754&+2.749,993&9945&-1.282&20&-1.698,180&18&+2.272& 10\v
-1.025&5.313,211,17&2818&+2.759,938&9943&-1.302&20&-1.698,162&19&+2.282& 10\\
-1.030&5.313,182,99&2883&+2.769,881&9943&-1.322&20&-1.698,143&19&+2.292& 10\\
-1.035&5.313,154,16&2949&+2.779,824&9941&-1.342&20&-1.698,124&20&+2.302& 10\\
-1.040&5.313,124,67&3018&+2.789,765&9939&-1.362&20&-1.698,104&20&+2.312& 10\\
-1.045&5.313,094,49&3087&+2.799,704&9939&-1.382&20&-1.698,084&20&+2.322& 10\v
-1.050&5.313,063,62&3159&+2.809,643&9936&-1.402&20&-1.698,064&21&+2.332& 10\\
-1.055&5.313,032,03&3232&+2.819,579&9936&-1.422&20&-1.698,043&22&+2.342&  9\\
-1.060&5.312,999,71&3306&+2.829,515&9934&-1.442&19&-1.698,021&22&+2.351& 10\\
-1.065&5.312,966,65&3383&+2.839,449&9932&-1.461&20&-1.697,999&22&+2.361& 10\\
-1.070&5.312,932,82&3460&+2.849,381&9931&-1.481&20&-1.697,977&23&+2.371& 10\v
-1.075&5.312,898,22&3541&+2.859,312&9929&-1.501&20&-1.697,954&24&+2.381& 10\\
-1.080&5.312,862,81&3623&+2.869,241&9928&-1.521&20&-1.697,930&24&+2.391& 10\\
-1.085&5.312,826,58&3706&+2.879,169&9926&-1.541&20&-1.697,906&25&+2.401& 10\\
-1.090&5.312,789,52&3791&+2.889,095&9924&-1.561&20&-1.697,881&25&+2.411& 10\\
-1.095&5.312,751,61&3879&+2.899,019&9922&-1.581&19&-1.697,856&26&+2.421& 10\v
-1.100&5.312,712,82&    &+2.908,941&    &-1.600&  &-1.697,830&  &+2.431&   \\
[\extrarowheight]\hline
\end{tabular}
\end{center}
\end{table*}

\end{document}